\documentclass[a4paper,11pt]{article}
\usepackage{jheppub} 
\usepackage{lineno}
\usepackage{subcaption}


\title{\boldmath Hypershadows {of higher dimensional black} objects:\\ a case study of cohomogeneity-one {d=5} Myers-Perry}

\author{João P. A. Novo,}
\author{Pedro V.P. Cunha}
\author{and Carlos A. R. Herdeiro}

\affiliation{Departamento de Matem\'atica  da Universidade de Aveiro and
	Center for Research and Development in Mathematics and Applications (CIDMA),
	Campus de Santiago, 3810-193 Aveiro, Portugal}

\abstract{
            \emph{What does a black hole look like?} {In $1+3$ spacetime dimensions, the optical appearance of a black hole is a bidimensional region  in the observer's sky often called the black hole shadow, as supported by the EHT observations. In higher dimensions this question is more subtle and observational setup dependent. Previous studies considered the shadows of higher dimensional black holes to remain bidimensional. We argue that the latter should be regarded as a tomography of a higher dimensional structure, the \emph{hypershadow}, which would be the structure  ``seen" by higher dimensional observers. As a case study we consider  the cohomogeneity-one Myers-Perry black hole in $1+4$ dimensions, and compute its tridimensional hypershadow.}
}

\numberwithin{equation}{section}

\begin{document}
\maketitle
\flushbottom
\newpage

\section{Introduction}
\label{sec:Intro}

Black holes (BHs) are one of the most distinctive predictions of General Relativity (GR). Despite their relevance, they are simple objects in four spacetime dimensions: stationary vacuum BHs are fully characterised by their mass and angular momentum, as asserted by several uniqueness theorems. Moreover they are always {mode} stable and possess a spherical event horizon~\cite{Hawking:1971vc,Hawking:1973uf}.

Whether this simplicity is connected to the BHs or rather just an artefact of the dimensionality of spacetime is a question worth pursuing. To do so one has to consider GR in $d>4$, whose systematic study is harder than in $d=4$, mainly because key solution generating techniques, such as a useful extension of the Newman-Penrose formalism to higher dimensions, are either not available or cannot span the full solution spectrum \cite{Emparan:2008eg}. Despite this difficulty there are plenty of results that indicate a richer {solutions landscape} in $d>4$~\cite{Emparan:2008eg}: event horizons with different topologies~\cite{Emparan:2001wn} and dynamical instabilities in BH horizons~\cite{Gregory:1993vy} which may lead to naked {singularities}~\cite{Lehner:2010pn}. Such higher dimensional solutions have been extensively considered in the context of string theory, gauge/gravity duality or brane-world scenarios.

The {historically pioneered}~\cite{kaluza,Klein:1926tv} and {minimal extension of} GR to higher dimensions is to consider $d=5$. Already in this case one has the higher dimensional analogues of Schwarzschild and Kerr BHs, respectively the Tangherlini solution~\cite{Tangherlini:1963bw} and the Myers-Perry solution~\cite{Myers:1986un}, with both solutions having topologically spherical horizons. Remarkably, these solutions can also coexist with black rings~\cite{Emparan:2001wn} within vacuum GR, for the same mass and spin, despite the black ring's horizon being of a toroidal topology. As such, the celebrated Kerr BH uniqueness theorems, valid in $d=4$, do not have simple generalisations in $d=5$.

The preceding comments illustrate how our expectations from $d=4$ gravity can be spoiled just by adding one extra dimension, it is therefore interesting to consider also how different properties of four dimensional black objects carry over when increasing the dimensionality. A property which {merits better understanding} in higher dimensions, despite valuable previous efforts~\cite{Papnoi:2014aaa,Hertog:2019hfb}, is the shadow of BHs. In $d=4$ the BH shadow is well established in the literature~\cite{Bardeen:1973tla,Cunha:2018acu} and has raised to prominence due to the observations of the Event Horizon Telescope (EHT) collaboration. Thus, an understanding of the higher dimensional counterpart is quite timely, {even if more academic}.

As a starting point we consider the $d=5$ cohomogeneity-one Myers-Perry BH, whose horizon is topologically a sphere, and has the advantage of having a fully separable Hamilton-Jacobi equation \cite{Vasudevan:2004mr}. This BH can be thought of as the counterpart of the Kerr solution in $d=4$, whose shadow boundary has the topology of $S^1$, so by analogy one expects that the boundary of the shadow of the $d=5$ BH is topologically a sphere, and the shadow is a volume instead of a surface. This would be the natural shadow for some higher dimensional beings living in a 5D spacetime whose retina is a volume instead of a surface like ours. To distinguish between what is defined in the literature by shadow we dub these shadows of $d=5$ BHs as \emph{hypershadows}. We show precisely that the hypershadow of these $d=5$ BHs has a topologically $S^2$ boundary. This raises the interesting question of considering whether the hypershadow of the Emparan-Reall black ring~\cite{Emparan:2001wn} might be topologically a torus ($S^1\times S^2$). However this black ring spacetime in not fully integrable and a numerical approach, beyond the scope of this work, would be necessary.
{In this respect, it can be remarked} that there are {engineered} $d=4$ line elements describing toroidal event horizons~\cite{KLEIHAUS2019134892}, whose shadows have been analysed; the boundary of the latter is {(indeed)} not connected for some observation angles~\cite{Cunha:2024ajc}.

This paper is organised as follows: in section \ref{section2} the 5 dimensional Myers-Perry solution is introduced. There we also discuss the separability of the equations of motion for null geodesics. The section closes with the cohomogeneity-one case, which will be considered for the remainder of the paper. Section \ref{sec:SPOs} is devoted to a systematic study of the Spherical Photon Orbits of the cohomogeneity-one MP BH, {a 5 dimensional counterpart of the study performed in \cite{Teo:2003ltt} for the Kerr BH}. The main result of the paper is presented in Section \ref{sec:Hypershadows} where the concept of the hypershadow is introduced, {and illustrated by} the hypershadow cast by the cohomogeneity-one MP BH. Three appendices provide technical results including a no bound orbits theorem for null orbits in cohomegeneity one Myers-Perry BHs.

\section{The 5D Myers-Perry solution}
\label{section2}

In 1986 Myers and Perry discovered the higher dimensional analogues of the Kerr solution, for arbitrary $d>4$ dimensions. These solutions have topologically spherical horizons and possess {$[(d-1)/2]$} independent angular momenta associated with the same number of orthogonal planes \cite{Myers:1986un}.
The 5D Myers-Perry solution is given by the line element:
\begin{align}
	\mathrm{d}s^2=&-\mathrm{d}t^2 + \left(x+a^2\right)\sin^2\theta\mathrm{d}\phi^2 + \left(x+b^2\right)\cos^2\theta\mathrm{d}\psi^2\, \nonumber\\
	& \quad \frac{\mu^2}{\rho^2}\left(\mathrm{d}t + a\sin^2\theta\mathrm{d}\phi + b\cos^2\theta\mathrm{d}\psi \right)^2+\frac{\rho^2}{4\Delta}\mathrm{d}x^2+\rho^2\mathrm{d}\theta^2\,,\label{eq:LineMP}
\end{align}
with
\begin{equation}
	\rho^2=x+a^2\cos^2\theta+b^2\sin^2 \theta\,,\quad\Delta=\left(x+a^2\right)\left(x+b^2\right)-\mu^2 x\,. 
\end{equation}
 It describes a topologically spherical BH, of mass $M=3\pi\mu^2/(8G)$, which simultaneously rotates in two orthogonal planes with angular momenta:
 \begin{equation}
     J_a=\frac{2}{3}Ma\,,\quad J_b=\frac{2}{3}Mb\,,
 \end{equation}
 where $a$ and $b$ are the spin parameters. The radial coordinate $x$ is the square of the usual Boyer-Lindquist like coordinate $r$, i.e. $x=r^2$. At spatial infinity the spatial part of the metric reduces to that of flat space in {bipolar} coordinates, hence the angular coordinates have the following ranges, $\theta\in\left[0,\pi/2\right]\,,{\phi\in\left[0,2\pi\right[\,,\psi\in\left[0,2\pi\right[}$. The metric possesses two event horizons corresponding to the roots of $\Delta$, with the outermost root being
\begin{equation}
	x_H=\frac{\mu^2-a^2-b^2+\sqrt{\left(\mu^2-a^2-b^2\right)^2-4a^2b^2}}{2}\,.\label{eq:xhor}
\end{equation}
In order to have $x_H\in\mathbb{R}$, the spin is bounded by $\mu\geqslant |a| +|b|$. Since we are only interested in the spacetime outside the outermost {(event)} horizon, we restrict the analysis to $x>x_H$. The horizon is also a Killing horizon of the vector field $\chi=\partial_t-\Omega_a\partial_\phi-\Omega_b\partial_\psi$, with the angular velocities $\Omega_i=i/\left(x_H +i^2\right)\,,i= a,b$.

The line element (\ref{eq:LineMP}) possesses three linearly independent Killing vector fields, to which the coordinate system is adapted, namely $\partial_t\,,\partial_\phi$ and $\partial_\psi$. There is a conserved quantity of motion along geodesics associated with each Killing vector, $p_t=-E\,, p_\phi=\Phi$ and $p_\psi=\Psi$, respectively. There exists an additional (non-trivial) constant of motion, $\mathcal{K}$, \emph{à la} Carter~\cite{Carter:1968rr}. The existence of this extra constant of motion implies that the geodesic motion on this background is fully integrable. The null geodesic equations on the background of the line element (\ref{eq:LineMP}) are known in the literature~\cite{Frolov:2003en, Diemer:2014lba} and can be analytically solved, as presented in~\cite{Diemer:2014lba}. However a complete discussion on the latter is beyond the scope of this work. Besides these Killing vector fields the metric is also invariant under the following transformation:
\begin{equation}
    a\leftrightarrow b\,,\quad \theta\leftrightarrow \frac{\pi}{2}-\theta\,,\quad \phi\leftrightarrow\psi\,.
\end{equation}

\subsection{Cohomogeneity-one and null geodesics}

Although the geodesic motion in the Myers-Perry spacetime is fully integrable, the equations are still fairly complex and depend on many parameters. Hence, as a case study the simpler case of the cohomogeneity-one solution, with $a=b$, will be considered. This solution possesses enhanced symmetries  \cite{Myers:2011yc}, and thus the equations of motion simplify greatly.

For the sake of clarity we write here the line element for the cohomogeneity-one solution:
\begin{align}
	\mathrm{d}s^2=&-\mathrm{d}t^2 + \left(x+a^2\right)\left(\frac{\mathrm{d}x^2}{4\Delta}+\mathrm{d}\theta^2+ \sin^2\theta\mathrm{d}\phi^2 +\cos^2\theta\mathrm{d}\psi^2\right)\, \nonumber\\
	& \quad + \frac{\mu^2}{\rho^2}\left[\mathrm{d}t + a\left(\sin^2\theta\mathrm{d}\phi + \cos^2\theta\mathrm{d}\psi\right) \right]^2\,.\label{eq:LineMPCoho}
\end{align}

The BH spin bound condition then becomes $|a| \leqslant \mu/2$. The null geodesic equations with $a=b$ reduce to~\cite{Frolov:2003en, Diemer:2014lba}:

\begin{align}
	\rho^{2}\dot{t}&=E\left(x+a^2\right)+\mu^2\frac{\left(x+a^{2}\right)^2}{\Delta}\left(E+\frac{a}{x+a^{2}}\left(\Phi+\Psi\right)\right)\,,\nonumber\\
	\rho^{2}\dot{\phi}&=\frac{\Phi}{\sin^{2}\theta}- \mu^2 a\frac{x+a^{2}}{\Delta}\left(E+\frac{a}{x+a^{2}}\left(\Phi+\Psi\right)\right)\,, \nonumber\\
	\rho^{2}\dot{\psi}&=\frac{\Psi}{\cos^{2}\theta}- \mu^2 a\frac{x+a^{2}}{\Delta}\left(E+\frac{a}{x+a^{2}}\left(\Phi+\Psi\right)\right)\,,\\
	\rho^{4}\dot{x}^2&={4\Delta\left(xE^2-\mathcal{K}\right)+ 4\mu^2\left(x+a^{2}\right)^2\left(E+\frac{a}{x+a^{2}}\left(\Phi+\Psi\right)\right)^{2}\,}, \nonumber\\
	\rho^{4}\dot{\theta}^2&=\mathcal{K}+E^2a^{2}-\frac{\Phi^{2}}{\sin^{2}\theta}-\frac{\Psi^{2}}{\cos^{2}\theta}\,. \nonumber
\end{align}

It will be useful to introduce a redefined Carter's constant $Q=E^2a^2+\mathcal{K}$. From the equation for $\dot{\theta}$ it then follows that
\begin{equation}
	{Q=\rho^4\dot{\theta}^2+\frac{\Phi^2}{\sin^2\theta}+\frac{\Psi^2}{\cos^2\theta}\,},\label{eq:CarterQ}
\end{equation}
thus $Q\geqslant0$. Moreover, one notices that the combination $\Phi+\Psi$ appears several times in the equations, which motivates the introduction of the quantities $\alpha=\Phi+\Psi$ and $\beta=\Phi-\Psi$. To further simplify the analysis, we can set the mass scale $\mu=1$, as well as $E=1$, with the latter corresponding to a redefinition of the affine parameter along null geodesics. The equations of motion then yield:
\begin{align}
	\rho^{2}\dot{t}&=\left(a^2+x\right) \frac{\left(a^2+x\right)^2+a^2+\alpha  a}{\left(a^2+x\right)^2-x}\,, \nonumber\\
	\rho^{2}\dot{\phi}&={\frac{\alpha +\beta}{2\sin^2\theta}  - a \frac{a^2+\alpha  a+x}{\left(a^2+x\right)^2-x}\,},\nonumber\\
	\rho^{2}\dot{\psi}&={\frac{\alpha -\beta}{2\cos^2\theta} - a\frac{a^2+\alpha  a+x}{\left(a^2+x\right)^2-x}\,}, \label{eq:CohoEqsMotion}\\
	{\frac{1}{4}\rho^{4}\dot{x}^2}&=\mathcal{X}={\Big(a^2-Q+x\Big)\Big[\left(a^2+x\right)^2-x\Big]+\Big[a(a+\alpha )+x\Big]^2\,},\nonumber\\
	\rho^{4}\dot{\theta}^2&={Q-\frac{(\alpha^2+\beta^2) + 2\alpha\beta \cos(2\theta)}{\sin^2(2\theta)}\,}.\nonumber
\end{align}
The full analytical solutions for these equations are presented in~\cite{Kagramanova:2012hw}. For the sake of this work, however, we are only interested in null geodesics with a constant radial coordinate, which are discussed in more detail in the next section.

\section{Spherical photon orbits}
\label{sec:SPOs}

The shadow of the Kerr BH is determined by the set of photon orbits with constant radial coordinate, known as spherical photon orbits \cite{Teo:2003ltt}. As will be proved later the same happens in the Myers-Perry spacetime, so this section will be devoted to the study of these photon trajectories with constant $x$. This type of orbits in the general Myers-Perry BH were first discussed in \cite{Bugden:2018uya}; since we consider only the cohomogeneity-one solution the equations are simpler and a more detailed and systematic analysis is possible.


\subsection{Motion along $\theta$}

The spherical photon orbits have constant radial coordinate, but often exhibit motion along $\theta$.

To study this latitudinal motion it is useful to consider the new coordinate $u = \cos\left(2\theta\right)$, with $u\in\left[-1,1\right]$. Then noticing that $\dot{u} = -2\sin(2\theta)\dot{\theta}$, the equation of motion for $u$ is:
\begin{equation}
	\frac{\rho^4}{4}\dot{u}^2=-Qu^2-2u\alpha\beta + Q - \alpha^2 - \beta^2\,,
\end{equation}
which is a simple quadratic equation in $u$ if $Q\neq 0$. The trivial case $Q=0$ implies $\{\Psi=0,\Phi=0\}$, as can be seen by inspecting Eq. \eqref{eq:CarterQ}, and it is not relevant for spherical photon orbits. Now we define $\mathcal{U}=-Qu^2-2u\alpha\beta + Q - \alpha^2 - \beta^2$. Physical motion can only occur when $\mathcal{U}\geqslant0$. To see where this condition is satisfied we start by looking at the values of $\mathcal{U}$ at the boundaries of the variable $u$:
\begin{align}
	\mathcal{U}(-1)&=-\left(\alpha-\beta\right)^2=-4\Psi^2<0\,,\nonumber\\	\mathcal{U}(1)&=-\left(\alpha+\beta\right)^2=-4\Phi^2<0\,.
\end{align}
Since $\mathcal{U}$ is negative at both ends and since $Q> 0$ as previously mentioned, the function $\mathcal{U}$ has a maximum. In order to have physical motion for $u\in\left[-1,1\right]$ this maximum must be non-negative. Such maximum occurs at
\begin{equation}
	\mathcal{U}^\prime=0\Leftrightarrow -2Qu_m-2\alpha\beta=0\Leftrightarrow u_m =-\frac{\alpha\beta}{Q}\,.
\end{equation}
At this point:
\begin{equation}
	\mathcal{U}(u_m)=Q-\alpha^2-\beta^2+\frac{\alpha^2\beta^2}{Q}\,.
\end{equation}
So, the parameters must satisfy:
\begin{equation}
	\begin{cases}
		\mathcal{U}\left(u_{m}\right)\geqslant 0\\
		-1\leqslant u_{m}\leqslant1
	\end{cases}\Leftrightarrow\begin{cases}
		Q-\alpha^{2}-\beta^{2}+\frac{\alpha^{2}\beta^{2}}{Q}\geqslant0\\
		-1\leqslant {-\frac{\alpha\beta}{Q}}\leqslant 1
	\end{cases}\Leftrightarrow\begin{cases}
		Q\geqslant\beta^{2} \label{eq:Qcondition}\\
		Q\geqslant\alpha^{2}
	\end{cases}\,,
\end{equation}
in order to have a region where $\mathcal{U}\geqslant0$. The last step in~\eqref{eq:Qcondition} is non-trivial, and the details can be found in Appendix~\ref{app:ProofQab}. Notice that when $Q=\alpha^2$ or when $Q=\beta^2$ one has $\mathcal{U}(u_m)=0$, so the orbit has constant $\theta$, and we are in the presence of a light ring. A light ring is to be interpreted as a geodesic whose tangent vector is a linear combination of the Killing vector fields \cite{Cunha:2017qtt}, hence it has constant $\theta$ and $x$.

\subsection{Radial motion}

The radial motion is determined by $\rho^{4}\dot{x}^2= 4\mathcal{X}$, see Eq.~\eqref{eq:CohoEqsMotion}. Spherical photon orbits have constant radial coordinate $x$, therefore they satisfy $\mathcal{X}=0$ and $\frac{d\mathcal{X}}{dx}=0$ simultaneously. The usefulness of the introduction of $\alpha$ now becomes apparent, since this system can be solved for the parameters $Q$ and $\alpha$ along these orbits, resulting in the following two branches:
\begin{align}
	Q_\pm=& {x+a^2+\frac{\left(x+a^2\right)^2-x}{\left(1\pm\sqrt{2}\sqrt{x+a^2}\right)^2}}\,,\label{eq:QSPO}\\
	\alpha_\pm =&-a-\frac{1}{a}\left[x+\frac{\left(a^2+x\right)^2-x}{1\pm\sqrt{2} \sqrt{a^2+x}}\right]\,.\label{eq:aSPO}
\end{align}

\begin{figure}
    \centering
    \includegraphics[width=0.6\linewidth]{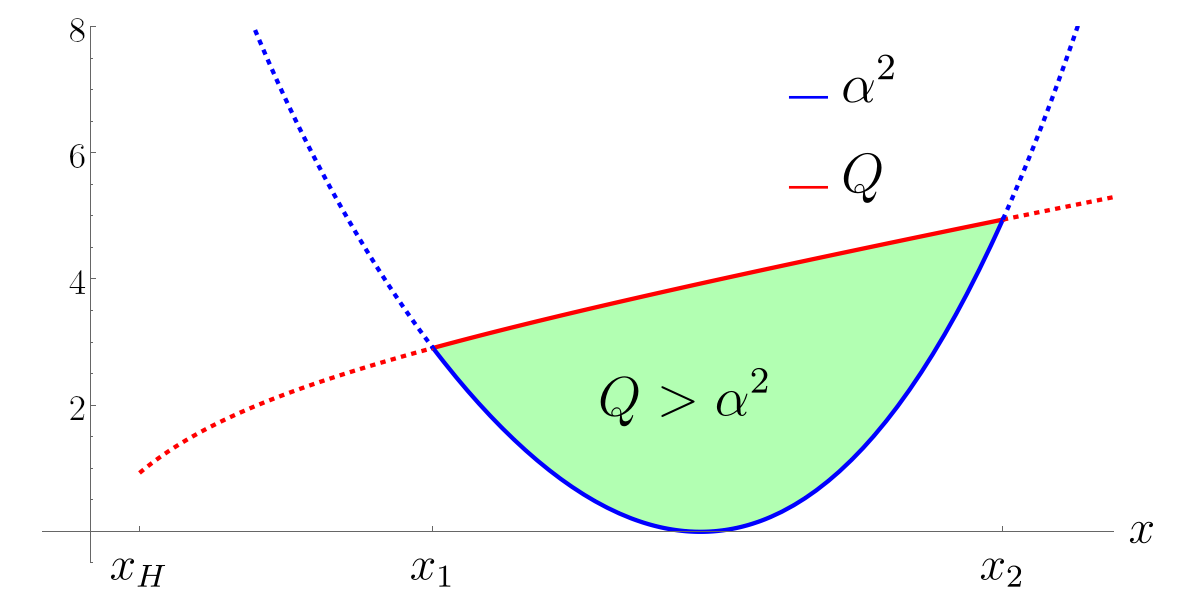}
    \caption{Representative plot of the functions $\alpha^2$ and $Q$ defined in Eqs. (\ref{eq:SPO1}) and (\ref{eq:SPO2}), with $a=1/4$. The region in the parameter space where spherical photon orbits are possible ($Q\geq\alpha^2$) is highlighted in green. This region is bounded by the radial coordinates $x_1$ and $x_2$.}
    \label{fig:AlphaQ}
\end{figure}

Latitudinal motion requires $Q-\alpha^2\geqslant 0$, so one has to check if both branches satisfy this condition. Then, for each branch one obtains:
\begin{align}
	Q_+-\alpha_+^2=-\frac{\left(x+a^2\right)^2}{a^2\left(1+\sqrt{2}\sqrt{x+a^2}\right)^2}\left[\left(a^2+x\right)^2+2 \sqrt{2} \left(a^2+x\right)^{3/2}+a^2+2 x\right]\,,\label{eq:plusBranch}\\
	Q_--\alpha_-^2={-\frac{\left(x+a^2\right)^2}{a^2\left(1-\sqrt{2}\sqrt{x+a^2}\right)^2}\left[\left(a^2+x\right)^2-2 \sqrt{2} \left(a^2+x\right)^{3/2}+a^2+2 x\right]}\,.\label{eq:minusBranch}
\end{align}
For the plus branch~\eqref{eq:plusBranch} one notices that the overall minus sign multiplies a factor which is a square, which is strictly positive since $x>x_H>0$, and a term in square brackets which is a sum of all positive terms, hence is also strictly positive. Thus $Q_+-\alpha_+^2<0$ and it is not a physical solution. On the other branch,~\eqref{eq:minusBranch}, the terms in square brackets do not have all the same sign, meaning that is possible to have $Q_--\alpha_-^2 \geqslant 0$. The negative branch is therefore the physical solution for spherical photon orbits:

\begin{align}
	Q(x)=&\, x+a^2+\frac{\left(x+a^2\right)^2-x}{\left(1-\sqrt{2}\sqrt{x+a^2}\right)^2}\,,\label{eq:SPO1}\\
	\alpha(x) =&-a-\frac{1}{a}\left[x+\frac{\left(a^2+x\right)^2-x}{1-\sqrt{2} \sqrt{a^2+x}}\right]\,.\label{eq:SPO2}
\end{align}
For the sake of simplicity, the minus sign of the physical branch was dropped. For reference, a representative plot of the condition $Q\geqslant \alpha^2$ is displayed in Fig.~\ref{fig:AlphaQ}.

\subsection{Photon region}

\begin{figure}
    \centering
    \includegraphics[width=0.7\linewidth]{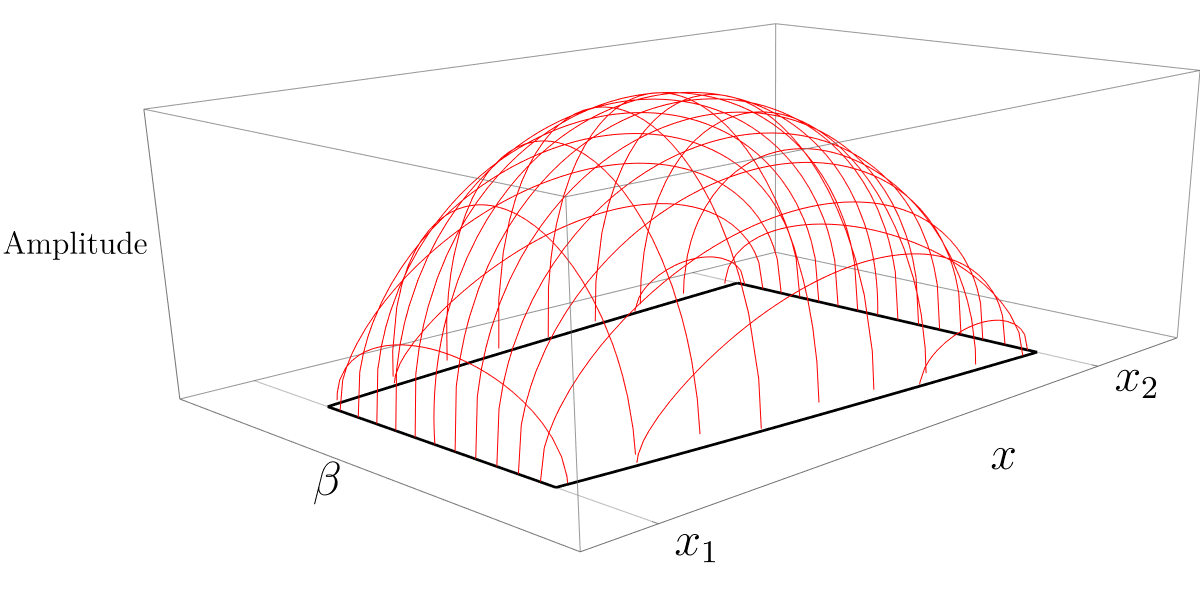}
    \caption{Illustrative representation of the photon region for $a=1/4$. The vertical axis represents a measure of the amplitude of motion along the latitude of the spherical photon orbits. In this plot it is clear that SPOs are only possible for radii between $x_1$ and $x_2$. This plot also illustrates the different possible light rings, which are located along the straight black lines with vanishing amplitude. Light rings with constant $x=x_{1,2}$ satisfy $Q=\alpha^2$, while those along the black lines with non-constant $x$ satisfy $Q=\beta^2$ instead.}
    \label{fig:photon-region}
\end{figure}

The physical domain of the spherical photon orbits, known as the {\it photon region}, is bounded by the condition $Q =\alpha^2$, which is satisfied at the radial coordinates $x^\pm$:
\begin{equation}
	x^\pm=1-a(a\mp1)+\sqrt{1\pm 2a}\,.
\end{equation}
It is useful to define the minimum and maximum of $x^\pm$ as:
\begin{equation}
    x_1 = \min\left(x^+, x^-\right)\,,\qquad     x_2 = \max\left(x^+, x^-\right)\,.
\end{equation}
The photon region is then restricted to the range $x\in [x_1,x_2]$, within which $Q \geqslant \alpha^2$ and $Q>0$ are both satisfied. However, recalling condition~\eqref{eq:Qcondition}, one must satisfy $Q\geqslant \beta^2$ as well, in order to have physical motion in the $\theta$ coordinate. This essentially implies that for each $x \in [x_1,x_2]$, the values of $\alpha(x),Q(x)$ are determined by~\eqref{eq:SPO1}-\eqref{eq:SPO2}, while $\beta$ can be freely changed within the range $ -\sqrt{Q(x)} \leqslant \beta \leqslant \sqrt{Q(x)}$, with each value of $\beta$ defining an independent spherical photon orbit at that radial location. The angular amplitude of the orbit can be determined by imposing that $\mathcal{U}\geqslant0$. A representation of the photon region is displayed in Fig.~\ref{fig:photon-region}. \\

Since light rings exist whenever $Q=\alpha^2$ or when $Q=\beta^2$, there is always a light ring orbit at each radius in the interval $x\in [x_1,x_2]$, albeit at different $\theta$ values. Notice how this differs from the Kerr spacetime, where there are only two light rings, occurring in the equatorial plane.

\section{The hypershadow}
\label{sec:Hypershadows}
\subsection{Local observer basis}

An observer at any point in the spacetime is able to construct a local orthonormal basis for the spacetime \cite{Bardeen:1973tla}, which can be expressed in terms of the coordinate basis $\{\partial_t,\partial_x,\partial_\theta,\partial_\phi,\partial_\psi\}$. Since the $\{x,\theta\}$ form a diagonal block on the metric the vectors $\partial_x$ and $\partial_\theta$ are already orthogonal to the other basis vectors. Thus the local tetrad basis can be expressed in the following form:
\begin{align}
	\hat{e}_{(t)}&=A^t\partial_t+B^t\partial_\phi+C^t\partial_\psi\,,\\
	\hat{e}_{(\phi)}&=A^\phi\partial_t+B^\phi\partial_\phi+C^\phi\partial_\psi\,,\\
	\hat{e}_{(\psi)}&=A^\psi\partial_t+B^\psi\partial_\phi+C^\psi\partial_\psi\,,\\
	\hat{e}_{(x)}&=A^x\partial_x\,,\\
	\hat{e}_{(\theta)}&=A^\theta\partial_\theta\,.	
\end{align}
Since the observer perceives the spacetime as locally Minkowski this basis should have the flat space normalisation, i.e.
\begin{equation}
	\hat{e}_{(\mu)}\cdot\hat{e}_{(\nu)}=	\hat{e}_{(\mu)}^\alpha g_{\alpha\beta}\hat{e}_{(\nu)} ^\beta=\eta_{(\mu)(\nu)}\,,
\end{equation}
where $\eta$ denotes the Minkowski metric. The coefficients $A^x$ and $A^\theta$ can then be straightforwardly computed:
\begin{align}
	A^x=&\frac{1}{\sqrt{g_{xx}}}=2\sqrt{\frac{(x+a^2)^2-x}{x+a^2}}\,,\\
	A^\theta=&\frac{1}{\sqrt{g_{\theta\theta}}}=\frac{1}{\sqrt{x+a^2}}\,.
\end{align}
Nine unknown coefficients remain with only six constraint equations. The three additional degrees of freedom are associated with rotations on the $\{t,\phi,\psi\}$ space, where each configuration yields an orthonormal set. Analogously, in Cartesian 3-space one can perform rotations with respect to each of the 3 axis and the axis remain orthogonal. Therefore, we fix this gauge freedom by imposing $A^\phi=A^\psi=B^\psi=0$. This choice is a 5 dimensional generalisation of a \emph{zero angular momentum observer} (ZAMO). Note however, that unlike the Kerr case there is a mixing of angular momenta with respect to infinity, as the physical momenta along $\phi$ is $p^{\left(\phi\right)}=B^\phi \Phi+C^\phi \Psi$.  With this choice we obtain:
\begin{align}
	A^t &= \sqrt{\frac{a^2+x}{\left(a^2+x\right)^2-x}+1} \,,\\
	B^t &= {-\frac{a}{\sqrt{\left(\left(a^2+x\right)^2-x\right) \left(\left(a^2+x\right)^2+a^2\right)}} }\,,\\
	C^t &= {-\frac{a}{\sqrt{\left(\left(a^2+x\right)^2-x\right) \left(\left(a^2+x\right)^2+a^2\right)}}}\,,\\
	B^\phi &={ \frac{1}{\sin\theta} \sqrt{\frac{a^2 \cos ^2\theta +\left(a^2+x\right)^2}{\left(a^2+x\right) \left( \left(a^2+x\right)^2+a^2\right)}}}\,,\\
	C^\phi &= {-\frac{ a^2 \sin\theta }{\sqrt{\left(a^2+x\right) \left(\left(a^2+x\right)^2+a^2\right) \left(a^2 \cos ^2\theta +\left(a^2+x\right)^2\right)}}} \,,\\
	C^\psi &= \frac{1}{\cos\theta} \sqrt{\frac{a^2+x}{a^2 \cos ^2\theta +\left(a^2+x\right)^2}} \,.
\end{align}
The signs were chosen in order to retrieve the standard orthogonal basis in bipolar coordinates at spatial infinity. Additionally it was required that $e_{\left(t\right)}\rightarrow \partial_t$.

The locally measured momenta are:
\begin{align}
	p^{(t)}&={-\hat{e}^\mu_{\left(t\right)} p_\mu=A^t E - B^t \Phi - C^t \Psi }\,, \nonumber\\ 
	p^{(\phi)}&=\hat{e}^\mu_{\left(\phi\right)} p_\mu=B^\phi \Phi + C^\phi \Psi\,,\nonumber\\ 
	p^{(\psi)}&=\hat{e}^\mu_{\left(\psi\right)} p_\mu=C^\psi \Psi\,, \label{eq:localMomenta} \\   
	p^{(x)}&=\hat{e}^\mu_{\left(x\right)} p_\mu=\frac{p_x}{\sqrt{g_{xx}}}\,, \nonumber\\
	p^{(\theta)}&=\hat{e}^\mu_{\left(\theta\right)} p_\mu=\frac{p_\theta}{\sqrt{g_{\theta\theta}}}\,. \nonumber
\end{align}
Note that a test particle with zero angular momenta ($\Phi=0=\Psi$) has no momentum along $\hat{e}_{\left(\phi\right)}$ or $\hat{e}_{\left(\psi\right)}$. This is characteristic of ZAMO frames, as the observer also has zero angular momentum with respect to infinity. However, a particle with $\Phi=0$ can have physical momentum along $\hat{e}_{\left(\phi\right)}$, if $\Psi\neq0$.

\subsection{Image space and impact parameters}

\begin{figure}
    \centering
    \includegraphics[width=0.99\linewidth]{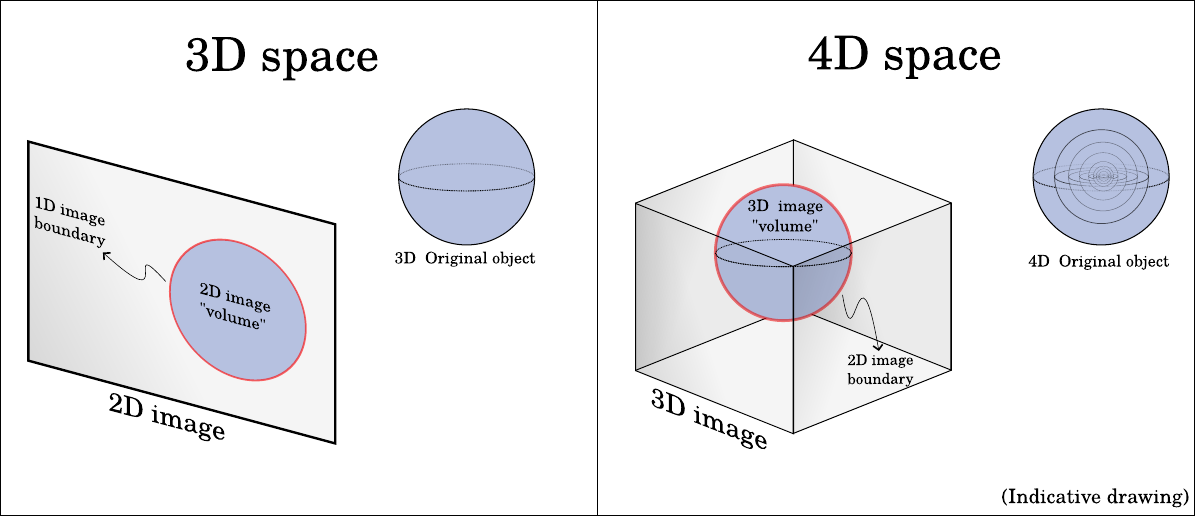}
    \caption{(Left panel): Taking the photograph of an object ($e.g.$ a sphere) in a spacetime with 3 spatial dimensions leads to a standard 2 dimensional image, with a 1 dimensional boundary. (Right panel): Taking the photograph of an object living in a spacetime with 4 spatial dimensions (5 in total when including the time coordinate), leads to a corresponding 3 dimensional image, with a 2 dimensional boundary ($i.e.$ a surface).}
    \label{fig:observer-setup}
\end{figure}

The shadow of a 5 dimensional BH have so far been presented and analysed in the literature as a 2D shape on an image plane at the observer's location~\cite{Papnoi:2014aaa,Hertog:2019hfb}. This setup is inspired by the way humans perceive images through the projection of light rays into our retina, which is a 2D surface. We argue however that, when considering the shadow of 5 dimensional objects, one should take the viewpoint of some higher dimensional beings whose retina equivalent is a volume, hence has 3 spatial dimensions, see Fig.~\ref{fig:observer-setup}. In such a case the shadow of a BH would be a volume delimited by some compact surface. Then the 2D shadow observed by humans would correspond to cross sections of this 3D shadow.

Therefore we construct a 3D image space with Cartesian coordinates $\{X,Y,Z\}$, which correspond to the impact parameters of the light rays, and are proportional to three observation angles $\{\Xi,\Lambda,\Upsilon\}$. We consider that the observer is placed at some point with coordinates $(x_o, \theta_o)$, assumed to be very far away from the BH, i.e. $x_o\gg\mu$.

The observation angles vanishes in the far away observation limit $x\rightarrow\infty$, following the same $\sim 1/r=1/\sqrt{x}$ decay law as in the four dimensional case, so the impact parameters are similarly defined as:
\begin{equation}
	X=-\sqrt{x_o}\,\Lambda\,,\quad Y=-\sqrt{x_o}\,\Upsilon\,,\quad Z=\sqrt{x_o}\,\Xi\,.
\end{equation}
Here the minus signs serve to make contact with the shadows of 4D BHs, wherein a point with $\Lambda>0$ should appear on the left of the image plane, i.e. $X<0$. This point is less evident in an 3D image space, but we decided to keep the minus signs.

The components of the physical momentum are constrained by $\left[p^{\left(t\right)}\right]^2=\mathsf{p}^2$ where $\mathsf{p}$ is the norm of the spatial part of the momentum, i.e. $\mathsf{p}^2=\sum \left[p^{\left(A\right)}\right]^2$, $A=\{x,\theta,\phi,\psi\}$. Therefore the spatial components can be parameterised by the detection angles as:
\begin{align}
	p^{\left(x\right)}&=\mathsf{p}\cos\Xi\cos\Lambda\cos\Upsilon\,\\
	p^{\left(\theta\right)}&=\mathsf{p}\sin\Xi\,\\
	p^{\left(\phi\right)}&=\mathsf{p}\cos\Xi\sin\Lambda\,\\
	p^{\left(\psi\right)}&=\mathsf{p}\cos\Xi\cos\Lambda\sin\Upsilon\,.
\end{align}
In this definition the origin of the observation angles $\{\Xi,\Lambda,\Upsilon\}=\{0,0,0\}$ corresponds to a direction pointing away from the BH.
When the observer is very far away from the BH the observation angles are expected to be very small:
\begin{equation}
	p^{\left(\theta\right)} \simeq\mathsf{p}\Xi\,, \quad  	p^{\left(\phi\right)} \simeq\mathsf{p}\Lambda\,, \quad 	p^{\left(\psi\right)} \simeq\mathsf{p}\Upsilon\,.
\end{equation}
In natural units the locally the energy of the photon is measured to be $\varepsilon=\mathsf{p}$, but this physical energy is also given by $\varepsilon=p^{\left(t\right)}$. Therefore, the impact parameters can be expressed as:
\begin{equation}
	X = -\sqrt{x_o}\, \frac{p^{\left(\phi\right)}}{p^{\left(t\right)}}\,, \quad 
	Y   =-\sqrt{x_o}\, \frac{p^{\left(\psi\right)}}{p^{\left(t\right)}}\,, \quad 
	Z =\sqrt{x_o}\, \frac{p^{\left(\theta\right)}}{p^{\left(t\right)}}\,.
\end{equation}

By combining the latter with Eqs. (\ref{eq:CohoEqsMotion}) and \eqref{eq:localMomenta}, and considering an observer in the far-away limit ($x_o\rightarrow\infty$) one obtains:
\begin{align}
	X&= {-\frac{(\alpha+\beta)}{2\sin\theta_o}= -\frac{\Phi}{\sin\theta_o}}\,,\label{eq:xprime}\\
	Y&= {-\frac{(\alpha-\beta)}{2\cos\theta_o}= -\frac{\Psi}{\cos\theta_o}} \,.\label{eq:yprime}\\
	Z&=\pm \frac{1}{\sin(2\theta_o)}\sqrt{Q\sin^2(2\theta_o) -\left(\alpha ^2+\beta ^2\right) -2\alpha\beta\cos(2\theta_o) }\,\label{eq:zprime} \ .
\end{align} 

The above equations connect the 3D image coordinates $\{X,Y,Z\}$ and the constants of motion $\{Q,\alpha,\beta\}$. The next subsection will apply this result to obtain the shadow of the 5D cohomogeneity-one MP BH.


\subsection{The shadow of the 5D cohomogeneity-one MP BH}

Since BHs do not emit light (at least classically), a background light source is necessary to provide the contrast needed to observe the BH's \textit{silhouette}. Imagine a distant light source with an angular size much larger than that of the BH, with the latter positioned between the observer and the light source. In this scenario, some scattering light rays from the source will fall into the BH and never reach the observer, creating a dark region in the image space known as the BH shadow. The boundary of this shadow is formed by photons that can get the closest to the event horizon but still reach infinity. Remarkably, the  shape and size of the shadow are independent of the specific details of the light source.\\

\begin{figure}
    \centering
    \includegraphics[width=0.5\textwidth]{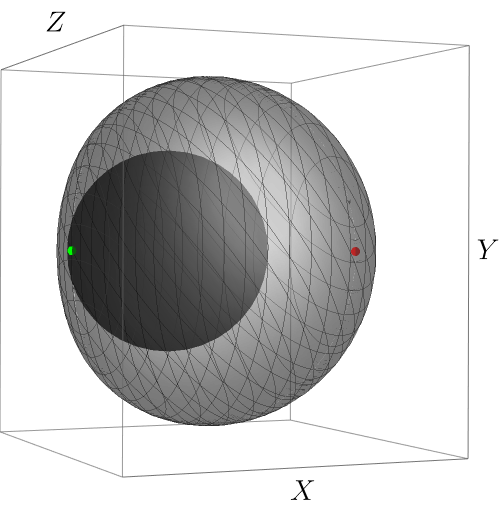}\includegraphics[width=0.5\textwidth]{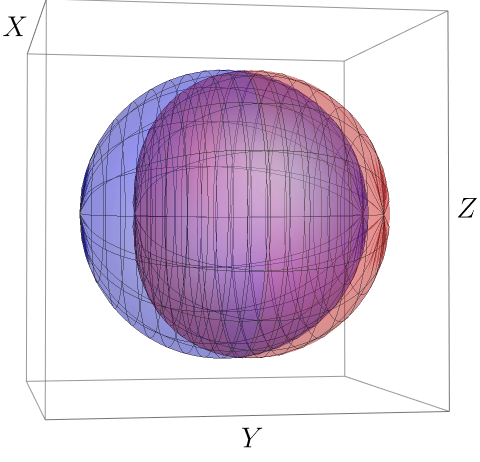}
    \caption{ (Left panel): The 2D boundary of the hypershadow of a cohomogeneity-one MP BH with spin $a=0.49$ and $\theta_o=\pi/2$. The green and red dots, $i.e.$ the light points, are determined by innermost and outermost light rings. We placed a reference sphere inside the hypershadow to highlight how it deviates from a sphere. (Right panel): The superimposed hypershadows of two MP BHs, in blue (red) with $a=0.1$  ($a=0.49$). }
    \label{fig:Hyper}
\end{figure}

The cohomogeneity-one MP spacetime satisfies a ``No-bound orbits theorem'' for null geodesics, as detailed in Appendix~\ref{app:NoBound}. This implies that in a scattering process, as described above, null geodesics can have at most two turning points in the radial coordinate, i.e. points where $\mathcal{X}=0$, outside the event horizon. By fixing the parameters $\{\Phi,\Psi,Q\}$ it is possible to choose the energy of the photon, $E$, to have two, one or zero turning points. The spherical photon orbits discussed in the previous section lie at the boundary between two or no turning points, and then represent the photons which can soar closer to the horizon and still reach the observer. Therefore the edge of the shadow is generated by these orbits, which are unstable, and thus contain photons which when lightly perturbed fall into or escape the BH. The 2D boundary of the hypershadow is thus obtained analytically by replacing into Eqs.~(\ref{eq:xprime})-(\ref{eq:zprime}) the values of $\alpha(x)$ and $Q(x)$ determined by Eqs.~\eqref{eq:SPO1}-\eqref{eq:SPO2} for the spherical photon orbits. The variation of both the radial coordinate in the range $x\in[x_1,x_2]$, as well as the parameter $\beta$ in the range $-\sqrt{Q(x)}\leqslant \beta \leqslant \sqrt{Q(x)}$ defines parametrically the 2D hypershadow surface edge (see Fig.~\ref{fig:Hyper}).\\

The effect of the spin parameter can be further appreciated in the right panel of Fig.~\ref{fig:Hyper}, which represents two hypershadows with different spins in the same plot. There is a spin squashing effect that shares a familiar resemblance to the  shadow deformation of the 4D Kerr BH as its spin increases. A notable difference with respect to Kerr is that for the MP hypershadow, it is always possible to see an image of four light rings: two of them correspond to the light rings at $x_{1,2}$ (highlighted as the red and green points in the left panel of Fig.~\ref{fig:Hyper}). Since these highlighted points will be referenced multiple times, let us simply designate them as {\it light points}, $i.e.$ the images of the light rings at $x_{1,2}$ with $Q=\alpha^2$. Distinct from the light points, there exist two additional images of light rings satisfying the condition $Q=\beta^2$, regardless of the observation angle $\theta_o$. However, these leave no noticeable signature in the hypershadow. In contrast to the MP case, the image of light rings on the Kerr shadow edge is only visible for observations in the equatorial plane.\\

\begin{figure}
    \centering
    \includegraphics[width=0.6\textwidth]{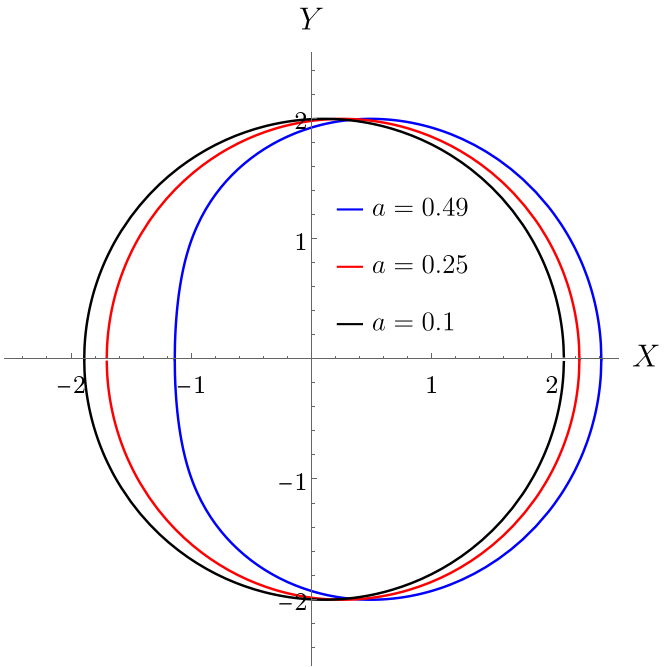}
    \caption{The hypershadow generating curves of several MP BHs due to the revolution symmetry. These images correspond to a $Z=0$ slice of the hypershadows with $\theta_o=\pi/2$.}
    \label{fig:Cross}
\end{figure}

Curiously, the shape of the MP hypershadow is independent of the observation angle $\theta_o$. Different values of $\theta_o$ amount to a simple rotation of the hypershadow in the image space. In addition, and perhaps even more surprisingly, the 2D boundary of the hypershadow  is a surface of revolution with respect to the axis that joins the light points (this can be thought of as a continuous version of the $\mathbb{Z}_2$ symmetry exhibited by the Kerr shadow). These symmetries can be explored to convey a much simpler parameterization of the MP hypershadow:
\begin{align}
    &X = -\alpha\sin\theta_o - \cos\upsilon \cos\theta_o\,\sqrt{Q-\alpha^2}\,,\\
    &Y =-\alpha\cos\theta_o + \cos\upsilon \sin\theta_o\,\sqrt{Q-\alpha^2}\,,\\
    &Z = \sin\upsilon \,\sqrt{Q-\alpha^2}\,,
    \label{eq:NewParameterization}
\end{align}
where $x\in [x_1,x_2]$ and $\upsilon\in [0,2\pi]$. As before, the values of $\alpha(x)$ and $Q(x)$ are determined by Eqs.~\eqref{eq:SPO1}-\eqref{eq:SPO2} for the spherical photon orbits.
Since these last few remarks are not immediately obvious from a simple inspection of Eqs.~(\ref{eq:xprime})-(\ref{eq:zprime}), a more detailed analysis can be found in Appendix~\ref{app:shadowsymmetries}. For reference, some hypershadow generating curves are displayed in Fig.~\ref{fig:Cross}, which appear remarkably Kerr-like.\\


As a final remark, cross sections of the hypershadow would correspond to the shadows of $5$ dimensional black object described in \cite{Hertog:2019hfb}.


\section{Conclusions and remarks}
\label{sec:Conclusions}

The optical appearance of four dimensional BHs have been discussed for a long time \cite{Bardeen:1973tla}, and are well established \cite{Cunha:2018acu}. They provide an observational avenue which rose to prominence with the observations of the EHT.\\

In an endeavour driven by the idea that considering how properties of GR vary for when the dimensionality of the spacetime is also varied, this paper set out to discuss and establish how would the ``shadow'' of a five dimensional BH would look like. Crucial to answer this question is the definition of the observational setup. Since spacetimes with one extra dimension were being considered, the observers should also have one extra dimension. This sets this work apart from others which studied shadows of $5$ dimensional objects \cite{Papnoi:2014aaa, Hertog:2019hfb}. So we considered the shadow as it would be perceived by a three dimensional observational device (like a retina), this means that the ``shadow'' is also a three dimensional object, a volume. To distinguish this volumetric ``shadow'' from the usual shadow, we dubbed the former \emph{hypershadow}. Moreover, we feel this choice is justified because the hypershadow, defined in this way, contains all the possible shadows a 2 dimensional observation device would see, which together would form a tomography of the hypershadow.\\

With the observational setup properly defined we computed and reported for the first time the hypershadow of a $5$-dimensional BH. For this purpose we considered the simplest rotating solution with integrable motion, the cohomogeneity-one Myers-Perry BH. As a byproduct of this study we also established a theorem which states that there are no bound null geodesics in the $5$-dimensional Myers-Perry spacetimes (in the cohomogeneity-one case), in the likeness of the one established by Wilkins for the Kerr solution \cite{Wilkins:1972rs}. A systematic study of the Spherical Photon Orbits for the cohomogeneity-one solution was also performed, where it was possible to realise that this solution possesses continuous families of light rings. As a consequence, the image of the light rings (the light points) are always visible in the hypershadow, regardless of the observation angle. Curiously, the shape and size of the hypershadow is also independent of the observational angle, which is qualitatively different from Kerr. However, the hypershadow of MP behaves in an akin way to the Kerr shadow under the influence of spin, with the hypershadow being squashed in one of the sides. Another non-trivial result is that the discrete $\mathbb{Z}_2$ symmetry exhibited by the Kerr shadow seems to be promoted in the cohomogeneity-one MP to a continuous symmetry, with its hypershadow becoming a surface of revolution with respect to the axis that connects the two light points. One can speculate whether this continuous symmetry is a consequence of having an MP BH with equal spins and thus with an enhanced symmetry. This could be checked by analysing the non-equal spins case.

\acknowledgments  
This work is supported by the Center for Research and Development in Mathematics and Applications
(CIDMA) through the Portuguese Foundation for Science and Technology (Fundação para a Ciência e a
Tecnologia), UIDB/04106/2020, UIDP/04106/2020, https://doi.org/10.54499/UIDB/04106/2020, https://doi.org/10.54499/UIDP/04106/2020. 
The authors also acknowledge support from the two projects PTDC/FISAST/3041/2020, http://doi.org/10.54499/PTDC/FISAST/3041/2020 as well as 2022.04560.PTDC, with doi https://doi.org/10.54499/2022.04560.PTDC.
This work has further been supported by the European
Horizon Europe staff exchange (SE) programme HORIZON-MSCA2021-SE-01 Grant No. NewFunFiCO101086251. P.C. is supported by the Individual CEEC program
http://doi.org/10.54499/2020.01411.CEECIND/CP1589/CT0035 of 2020, funded by the Portuguese Foundation for Science and Technology (Fundação para a Ciência e a Tecnologia). J.N. is supported by the FCT grant 2021.06539.BD.

\appendix
\section{Proof that $Q\geqslant \alpha^2$ and $Q\geqslant \beta^2$}
\label{app:ProofQab}

Recalling the discussion in Section~\ref{sec:SPOs}, the condition for real motion in the $\theta$ coordinate is translated into the two conditions:
\begin{equation}
\label{eq:appendixA}
f(Q) \equiv Q^2 -Q(\alpha^2+\beta^2) + \alpha^2\beta^2 \geqslant 0 \qquad\land\qquad Q\geqslant \alpha\beta \geqslant -Q\,,
\end{equation}
with $Q>0$.\\

It is useful to obtain the points at which $f(Q)=0$, which yields two roots~\footnote{These roots are different that the quantities~\eqref{eq:QSPO}, althought the notation is similar.}:
\[Q^\pm = \frac{\alpha^2+\beta^2 \pm |\alpha^2-\beta^2| }{2}\,.\]

We have two cases:
\begin{align}
\label{eq:AppendixoPtions1}
    &\textrm{If } \alpha^2\geqslant \beta^2 : \left\{
  \begin{array}{lr}
    Q^+ = \alpha^2\geqslant 0 &\\
    Q^- = \beta^2\geqslant 0
  \end{array}
\right.\\
\label{eq:AppendixoPtions2}
&\textrm{If } \beta^2\geqslant \alpha^2 : \left\{
  \begin{array}{lr}
    Q^+ = \beta^2\geqslant 0 &\\
    Q^- = \alpha^2\geqslant 0
  \end{array}
\right. \ .
\end{align}

Next, it is useful to introduce the quantity $\widetilde{Q}\equiv \sqrt{Q^+Q^-}=|\alpha \beta|$.
Then one can show the following relations:
\begin{align*}
    Q^+\geqslant Q^- & \implies Q^+Q^-\geqslant \left({Q^-}\right)^2 \implies \sqrt{Q^+Q^-}\equiv \widetilde{Q}\geqslant Q^-\\
    Q^+\geqslant Q^- & \implies \left({Q^+}\right)^2\geqslant Q^+Q^-  \implies Q^+\geqslant  \widetilde{Q}\,,
\end{align*}
hence $Q^+ \geqslant \widetilde{Q} \geqslant Q^-$. In addition, from conditions~\eqref{eq:appendixA} we have:
\[Q\geqslant \alpha\beta \geqslant -Q \iff Q \geqslant |\alpha\beta|=\widetilde{Q}\,.\]
Given the fact that $f(Q)$ is a quadratic function with positive concavity, this implies that the necessary and sufficient condition to satisfy~\eqref{eq:appendixA} is $Q \geqslant Q^+$. Given the options~\eqref{eq:AppendixoPtions1}-\eqref{eq:AppendixoPtions2}, this concludes the proof:
\[Q\geqslant \alpha^2\,\quad\land\quad Q\geqslant \beta^2\,.\]
\[\]

\section{``No-bound theorem'' - Cohomogeneity-one Myers-Perry spacetime}
\label{app:NoBound}

In the main text, it was assumed that there are no bound null orbits in the cohomogeneity-one Myers-Perry BH, and as such the hypershadow edge was determined by the spherical photon orbits.
In this appendix we shall indeed show that bound orbits for null geodesics are not possible outside the event horizon. The equations governing the radial motion of null geodesics are repeated for convenience~\cite{Frolov:2003en, Diemer:2014lba}:
\begin{align}
	&\rho^{4}\dot{x}^2=4\mathcal{X}\,,\label{eq:appendxiEqX} \\
	&\mathcal{X}={\Delta\left[xE^{2} -\mathcal{K}\right]} \,+ \, \mu^{2}\left(x+a^{2}\right)^2\left(E+\frac{a}{x+a^{2}}(\Phi+\Psi)\right)^{2}\nonumber
\end{align}
It will be useful to work with the following parameters:
\begin{align}
        \mathcal{Q}&=\mathcal{K}+2E^2 a^2\,,\qquad\quad \mu_a^2=\mu^2-4a^2\,,\\
	\alpha&=\Phi+\Psi\,,\qquad\qquad\,\,\,\, \beta=\Phi-\Psi\,,
\end{align}
In addition, one can introduce a new radial coordinate $y$, defined by:
\begin{equation}
y = 2x - \frac{\mu_a^2+\mu^2}{2}\,.
\end{equation}

This new radial coordinate $y$ is always positive outside the horizon ($y > 0$), since in terms of the new parameters the horizon coordinate $x_H$ becomes:
\begin{equation}
	x_H =\frac{\mu_a^2+\mu^2}{4} +\frac{1}{2}\sqrt{\mu_a^2 \mu^2}\,\,
\end{equation}
which implies
\begin{equation}
    x>x_H \implies \frac{y}{2} + \frac{\mu_a^2+\mu^2}{4} > x_H \implies y >\sqrt{\mu_a^2 \mu^2} \geqslant 0\,.
\end{equation}
We remark\footnote{The spin constrain $\mu \geqslant 2|a|$ implies $\mu^2 \geqslant 4a^2$, which implies $\mu_a^2\geqslant 0$.} that $\mu^2_a \geqslant 0$ if an event horizon exists.\\

Curiously, one can write the function $4\mathcal{X}$ in~\eqref{eq:appendxiEqX} as a cubic polynomial, in terms of the new radial coordinate $y$:
\begin{equation}
	4\mathcal{X}=\frac{E^2}{2}y^3 + c_2 y^2+c_1 y + c_0 \,.\label{eq:yPoly}
\end{equation}
Where
\begin{align}
	c_2&=\left(\frac{ 3 \mu ^2+ 2a^2}{2}\right)E^2 - \mathcal{Q}\,,\\
	c_1&=\frac{E^2\,\mu^2}{2} \left(3 \mu^2+ 4a^2 \right)+ 4 a\,\alpha\mu^2\,E\,,\\
	c_0&= \mathcal{Q}\,\mu_a^2\mu^2 + 4a^2\alpha^2\mu^2 +  E^2\mu^2 \Big(4a^4 +\mu^2a^2 +\frac{\mu^4}{2} \Big) + 4a\,\alpha\mu^4\,E\,.
 \label{eq:c2c1c0}
\end{align}

At this point we shall divide the proof into four special cases, as detailed below.
\begin{itemize}
    \item[i)] generic case with $E\neq 0$, and $0<|a|<\mu/2$.
    \item[ii)] $a=0$ (no spin);
    \item[iii)] $|a|=\mu/2$ (extremality);
    \item[iv)] $E=0$ (null geodesic with zero energy with respect to infinity);
\end{itemize}

\subsection{Generic case with $E\neq 0$, and $0<|a|<\mu/2$}
\label{subsection1}
A crucial observation is that in order to have a region where the null geodesic motion is bounded, then the function $\mathcal{X}$ must have at least three distinct roots outside the horizon, $i.e.$ with $y>0$. This will be proven bellow to be impossible, since $\mathcal{X}$ can have at most two such roots, and therefore bound null geodesics are not possible in the 5D Myers-Perry spacetime.\\

As previously discussed, the region outside the event horizon has always $y>0$, and so we are interested in determining the number of \emph{positive} roots of $\mathcal{X}$. To this end, one can use the following well-known mathematical result:

$ $\\
{\bf Descartes' rule of signs:}\\
{\it If the nonzero terms of a single-variable polynomial with real coefficients are ordered by descending variable exponent, then the number of positive roots of the polynomial is either: i) equal to the number of sign changes between consecutive (nonzero) coefficients, or ii) is less than it by an even number. In this context, a root of multiplicity $k$ is counted as $k$ roots.} (see $e.g.$ \cite{Anderson1998447})\\

Since in Eq. (\ref{eq:yPoly}) the highest coefficient is always positive ($E\neq 0$ by assumption), in order to have (at most) three roots the remaining coefficients must verify:
\begin{equation}
	c_2<0\,,\quad c_1>0\,,\quad c_0<0\,.
\end{equation}
This implies in particular that (taking dimensional analysis into consideration):
\begin{equation}
	\mu^2 c_1-(\mu^4 c_2+c_0)>0\,.
     \label{eq:inequalityrelation}
\end{equation}
After substitutions this yields:
\begin{equation}
    4a^2\mu^2\,\mathcal{Q} -4a^2\alpha^2\mu^2  - E^2 \left( 4a^4\mu^2 +\frac{\mu^6}{2}\right)\,\,>\,0
\end{equation}

Since the coefficient in $\mathcal{Q}$ is positive and non-zero (by assumption $a\neq 0$), it is possible to write:
\begin{equation}
	\mathcal{Q}>\frac{\left[\cdots\right]}{4a^2\mu^2}\,,\label{eq:ineq1}
\end{equation}
where the details in the numerator were suppressed as a shorthand notation. Similarly, from the condition $c_0<0$, and given that the $\mathcal{Q}$ coefficient is also positive and non-zero (by assumption $|a|\neq \mu/2$) we obtain:
\begin{equation}
	\mathcal{Q}<\frac{\left\{\cdots\right\}}{\mu_a^2\mu ^2}\,.\label{eq:ineq2}
\end{equation}

Now, we can put together Eqs. (\ref{eq:ineq1}) and (\ref{eq:ineq2}) to write:
\begin{align}
	\frac{\left[\cdots\right]}{4a^2\mu^2}-\frac{\left\{\cdots\right\}}{\mu_a^2\mu ^2}<0\nonumber\\
	\frac{f\!\left(E\right)}{8\mu^2\mu_a^2a^2} <0\,.
\end{align}
Where $f\left(E\right)=\gamma_2 E^2+\gamma_1 E +\gamma_0$ with:
\begin{align}
	\gamma_2 &= \mu^4(16a^4 + \mu^4) \,,\\
	\gamma_1 &= 32\alpha a^3\mu^4\,,\\
	\gamma_0 &= 8\alpha^2a^2\mu^4\,.
\end{align}
Since $\gamma_2>0$  we have that $f(E)$ is a convex parabola, hence $f(E)<0$ for $E\in\left[E_1,E_2\right]$, where $E_1\,,E_2$ are the roots of $f$. The existence of these roots can be asserted from the discriminant of the quadratic equation, which, after some algebraic simplifications, yields:
\begin{equation}
	\gamma_1^2-4\gamma_0\gamma_2= 32a^2\alpha^2\mu^8(4a^2+\mu^2)(4a^2-\mu^2)\,.
\end{equation}
All factors are positive, except $4a^2-\mu^2$ which is negative since the spin constrain implies $\mu_a^2>0 \implies \mu^2> 4a^2$. A negative discriminant in turn implies that the equation $f(E)=0$ has no real roots. Therefore Eqs. (\ref{eq:ineq1}) and (\ref{eq:ineq2}) are not compatible, and the equation $\mathcal{X}=0$ has at most two roots outside the horizon. Therefore, by the previous discussion one can assert that there are no bound null geodesics outside the event horizon in the subcase~\ref{subsection1}, with $E\neq 0$, and $0<|a|<\mu/2$ .\\

\subsection{Case $a=0$ (no spin)}
\label{subsection2}

If there is no spin, $a=0$, then relation~\eqref{eq:inequalityrelation} becomes:
\begin{equation}
    \mu^2c_1 - (\mu^4c_2 + c_0) >0 \iff  -\frac{1}{2}\mu^6 E^2 >0\,,
\end{equation}
which is not possible for real numbers, and thus the equation $\mathcal{X}=0$ has at most two roots outside the horizon. Therefore, there are no bound null geodesics outside the event horizon in the subcase~\ref{subsection2}.

\subsection{Case $|a|=\mu/2$ (extremality)}
\label{subsection3}

At extremality, with $a=\pm \mu/2$, the term $c_0$ in~\eqref{eq:c2c1c0} becomes:
\begin{equation}
    c_0 = \mu^4(\alpha \pm E \mu)^2\,.
    \label{eq:cofinal}
\end{equation}
Having three roots for $\mathcal{X}=0$ requires $c_0<0$. However, since~\eqref{eq:cofinal} is never negative, we conclude that there can be at most two roots with $y>0$. Therefore, there are no bound null geodesics outside the event horizon in the subcase~\ref{subsection3}, with $|a|=\mu/2$ (extremality).

\subsection{Case $E=0$}
\label{subsection0}

When $E=0$, the polynomial~\eqref{eq:yPoly} becomes quadratic in the $y$ radial coordinate:
\begin{equation}
    4\mathcal{X}= -\mathcal{Q}\,y^2 + \Big(4a^2\alpha^2\mu^2 + \mathcal{Q}\mu^2\mu_a^2\Big)\,.
\end{equation}
If $\mathcal{Q}\neq 0$, there can be at most one positive root, and as such its is not possible to have a region $y>0$ with bounded null geodesics.\\
If $\mathcal{Q}=0$ there might be a possibility of trivially bounded motion, since $\mathcal{X}=0$ for any positive $y$ if either $\alpha$ or $a$ are zero.  However, in such a scenario either the motion in the $\theta$ coordinate is not real ($\Theta<0$), or there will be no motion in the time coordinate ($\dot{t}=0$), which is unphysical.\\

Therefore, there are no bound null geodesics outside the event horizon in the subcase~\ref{subsection0}, with $E=0$.
\[\]
\subsection*{Conclusion}

Bringing together the results of the cases~\ref{subsection1},~\ref{subsection2},~\ref{subsection3},~\ref{subsection0}, this concludes the proof that bounded geodesic motion is not possible outside the horizon of a cohomogeneity-one Myers-Perry BH.

\[\]
\section{Non-trivial properties of the MP hypershadow}
\label{app:shadowsymmetries}

The light points of the hypershadow play an import role in understanding some of its non-trivial properties. The light points are the image of the innermost and outermost light rings, obtained by the condition $Q=\alpha^2$. Including this condition into~\eqref{eq:zprime} for the $Z$ coordinate leads to:
\[Z=\pm \frac{1}{\sin(2\theta_o)}\sqrt{-\Big(\alpha\cos(2\theta_o)+\beta\Big)^2}\,.\]
Since $Z\in \mathbb{R}$, this implies that $Z=0$ and $\beta=-\alpha\cos(2\theta_o)$. Inserting this condition into Eqs.~\eqref{eq:xprime} and~\eqref{eq:yprime} similarly leads to:\\
\begin{equation}
     X = -\alpha\,\sin\theta_o\,,\qquad Y = -\alpha\,\cos\theta_o\,.
\end{equation}
This implies that the light points always exist on the plane $Z=0$ and changing the observation angle $\theta_o$ amounts to a rotation of both light points around the coordinate origin.\\

As it will turn out, the axis connecting both light points is also a symmetry axis of the hypershadow. To show this point it will be helpful to introduce new image coordinates $\{X',Y',Z'\}$, with the new coordinate $X'$ adapted to the axis connecting both light points. The coordinate transformation between the old and the new coordinate set is simply provided by a rotation along the $Z$-axis:
\begin{equation}
\begin{pmatrix}
X'\\
Y'\\
Z'
\end{pmatrix}
=
\begin{pmatrix}
\phantom{-}\sin\theta_o & \cos\theta_o & 0\\
-\cos\theta_o & \sin\theta_o & 0 \\
0 & 0 & 1
\end{pmatrix}
\begin{pmatrix}
X\\
Y\\
Z
\end{pmatrix}
\,.
\label{eq:appendixTransform}
\end{equation}
Under these new image coordinates, Eqs.~\eqref{eq:xprime} and~\eqref{eq:yprime} become:
\begin{align*}
    X'&= -\alpha\,,\\
    Y'&= \frac{(\alpha+\beta)}{2\tan\theta_o} + \frac{\tan\theta_o}{2}\,(\beta-\alpha)\,.
\end{align*}
The latter is valid for a generic point of the hypershadow edge, not only the light points. By computing the radius $R$ of a cross-section of the hypershadow with fixed $X'$, one obtains:
\[R^2 = {Y'}^2 + {Z'}^2 = Q-\alpha^2\,.\]
Hence, the cross section radius $R=\sqrt{Q-\alpha^2}$ only depends on the radial coordinate $x$ of the spherical photon orbits. In particular, this cross-section radius is independent on both $\beta$ and $\theta_o$. Since $X'$ also only depends on the radial coordinate $x$, this implies:
\begin{itemize}
    \item The cross sections of the hypershadow with constant $X'$ are perfect circles, with their radius possibly changing along the $X'$-axis. The $X'$-axis is indeed a symmetry axis of the hypershadow.
    \item Changing $\theta_o$ simply amounts to a rotation of the entire hypershadow. This is a consequence of the $X'$-axis connecting both light points, and (as previously discussed) changing the observation angle $\theta_o$ amounts to a rotation of both light points. In addition the cross-section radius $R$ does not depend on $\theta_o$. As a consistency check, one can realise that in the original coordinates $X^2+Y^2+Z^2=Q$, which is also independent of the observation angle $\theta_o$.\\
\end{itemize}

The hypershadow can be parameterized in a simpler way using this non-trivial symmetry. Since cross-section with constant $X'$ are circles, one can write:
\begin{equation*}
    X'=-\alpha\,,\qquad Y'=R\cos\upsilon\,,\qquad Z'=R\sin\upsilon\,.
\end{equation*}
where $\upsilon\in[0,2\pi]$ is an auxiliary angle. By inverting the coordinate transformation~\eqref{eq:appendixTransform} one recovers the original equations~\eqref{eq:NewParameterization} reported in the main text.

\bibliographystyle{JHEP}
\bibliography{biblio}
\end{document}